\documentclass[11pt,preprintnumbers,titlepage,nofootinbib]{revtex4-2}
\usepackage[latin9]{inputenc}
\usepackage{amsmath}
\usepackage{amssymb}
\usepackage{graphicx}
\usepackage{wasysym}
\usepackage[unicode=true,
 bookmarks=false,
 breaklinks=false,pdfborder={0 0 1},backref=section,colorlinks=false]
 {hyperref}
\hypersetup{
 colorlinks,linkcolor=blue,citecolor=blue,urlcolor=blue}

\makeatletter

\providecommand{\tabularnewline}{\\}
\newcommand{\lyxdot}{.}

\usepackage{wasysym}
\usepackage{array}
\usepackage{multirow}
\usepackage{wasysym}
\usepackage{wasysym}
\usepackage{amsfonts}
\usepackage{subfigure}
\usepackage{cleveref}
\usepackage{listings}
\usepackage{xcolor}
\usepackage{array}
\usepackage{url}
\usepackage{color}

\providecommand{\tabularnewline}{\\}
\@ifundefined{textcolor}{}{
\definecolor{BLACK}{gray}{0}
\definecolor{WHITE}{gray}{1}
\definecolor{RED}{rgb}{1,0,0}
\definecolor{GREEN}{rgb}{0,1,0}
\definecolor{BLUE}{rgb}{0,0,1}
\definecolor{CYAN}{cmyk}{1,0,0,0}
\definecolor{MAGENTA}{cmyk}{0,1,0,0}
\definecolor{YELLOW}{cmyk}{0,0,1,0}
}

\makeatother

\begin{document}
\preprint{CTP-SCU/2023031}
\title{Gravitational Lensing by Transparent Janis-Newman-Winicour Naked Singularities}
\author{Deyou Chen$^{a,b}$}
\email{deyouchen@hotmail.com}

\author{Yiqian Chen$^{c}$}
\email{yqchen@stu.scu.edu.cn}

\author{Peng Wang$^{c}$}
\email{pengw@scu.edu.cn}

\author{Tianshu Wu$^{c}$}
\email{wutianshu@stu.scu.edu.cn}

\author{Houwen Wu$^{c,d}$}
\email{hw598@damtp.cam.ac.uk}

\affiliation{$^{a}$School of Science, Xihua University, Chengdu, 610039, China}
\affiliation{$^{b}$Key Laboratory of High Performance Scientific Computation,
Xihua University, Chengdu, 610039, China}
\affiliation{$^{c}$Center for Theoretical Physics, College of Physics, Sichuan
University, Chengdu, 610064, China}
\affiliation{$^{d}$Department of Applied Mathematics and Theoretical Physics,
University of Cambridge, Wilberforce Road, Cambridge, CB3 0WA, UK}
\begin{abstract}
The Janis-Newman-Winicour (JNW) spacetime can describe a naked singularity
with a photon sphere that smoothly transforms into a Schwarzschild
black hole. Our analysis reveals that photons, upon entering the photon
sphere, converge to the singularity in a finite coordinate time. Furthermore,
if the singularity is subjected to some regularization, these photons
can traverse the regularized singularity. Subsequently, we investigate
the gravitational lensing of distant sources and show that new images
emerge within the critical curve formed by light rays escaping from
the photon sphere. These newfound images offer a powerful tool for
the detection and study of JNW naked singularities.
\end{abstract}
\maketitle
\tableofcontents{}

\section{Introduction}

Gravitational lensing, which involves the bending of light in curved
space as predicted by general relativity \cite{Dyson:1920cwa,Einstein:1936llh,Eddington:1987tk},
is an intriguing and fundamental phenomenon with significant importance
in astrophysics and cosmology. Over the past decades, extensive research
has been dedicated to studying gravitational lensing, contributing
significantly to addressing crucial topics, including the distribution
of structures \cite{Mellier:1998pk,Bartelmann:1999yn,Heymans:2013fya},
dark matter \cite{Kaiser:1992ps,Clowe:2006eq,Atamurotov:2021hoq},
dark energy \cite{Biesiada:2006zf,Cao:2015qja,DES:2020ahh,DES:2021wwk},
quasars \cite{SDSS:2000jpb,Peng:2006ew,Oguri:2010ns,Yue:2021nwt}
and gravitational waves \cite{Seljak:2003pn,Diego:2021fyd,Finke:2021znb}.
In the context of an idealized lens model featuring a distant source
and a Schwarzschild black hole, the slight deflection of light in
a weak gravitational field results in the observation of primary and
secondary images. Moreover, strong gravitational lensing near the
photon sphere gives rise to an infinite sequence of higher-order images,
referred to as relativistic images, on both sides of the optic axis
\cite{Virbhadra:1999nm}. Notably, relativistic images exhibit minimal
sensitivity to the characteristics of the astronomical source, making
them valuable tools for exploring the nature of black hole spacetime.

The Event Horizon Telescope collaboration has recently achieved high
angular resolution \cite{Akiyama:2019cqa,Akiyama:2019brx,Akiyama:2019sww,Akiyama:2019bqs,Akiyama:2019fyp,Akiyama:2019eap,Akiyama:2021qum,Akiyama:2021tfw,EventHorizonTelescope:2022xnr,EventHorizonTelescope:2022vjs,EventHorizonTelescope:2022wok,EventHorizonTelescope:2022exc,EventHorizonTelescope:2022urf,EventHorizonTelescope:2022xqj},
allowing for the study of gravitational lensing in the strong gravity
regime and reigniting interest in the shadow of black hole images
and the associated phenomenon of strong gravitational lensing \cite{Falcke:1999pj,Claudel:2000yi,Eiroa:2002mk,Virbhadra:2008ws,Yumoto:2012kz,Wei:2013kza,Zakharov:2014lqa,Atamurotov:2015xfa,Dastan:2016bfy,Cunha:2016wzk,Wang:2017hjl,Amir:2017slq,Ovgun:2018tua,Perlick:2018iye,Zhu:2019ura,Bambi:2019tjh,Mishra:2019trb,Vagnozzi:2019apd,Kumar:2019pjp,Ma:2019ybz,Allahyari:2019jqz,Zeng:2020dco,Zeng:2020vsj,Roy:2020dyy,Li:2020drn,Kumar:2020hgm,Vagnozzi:2020quf,Khodadi:2020jij,Chowdhuri:2020ipb,Saurabh:2020zqg,Zhang:2020xub,Gan:2021xdl,Gan:2021pwu,Sarkar:2021djs,Guerrero:2022qkh,Virbhadra:2022iiy,Vagnozzi:2022moj,Guo:2022muy,Chen:2022qrw,Ghosh:2022mka,AbhishekChowdhuri:2023ekr}.
This research has shown a close relationship between strong gravitational
lensing and bound photon orbits, which lead to the existence of photon
spheres in spherically symmetric black holes. Interestingly, horizonless
ultra-compact objects have been discovered to possess photon spheres,
effectively mimicking black holes in various observational simulations
\cite{Schmidt:2008hc,Guzik:2009cm,Liao:2015uzb,Goulart:2017iko,Nascimento:2020ime,Qin:2020xzu,Islam:2021ful,Tsukamoto:2021caq,Junior:2021svb,Olmo:2021piq,Ghosh:2022mka}.
Future observations may offer the opportunity to test observable effects
characteristic of specific spacetimes, thus enabling a precise distinction
between black hole mimickers.

Among black hole mimickers, naked singularities have attracted considerable
attention. Although the cosmic censorship conjecture prohibits the
formation of naked singularities, they can arise through the gravitational
collapse of massive objects under specific initial conditions \cite{Shapiro:1991zza,Joshi:1993zg,Harada:1998cq,Joshi:2001xi,Goswami:2006ph,Banerjee:2017njk,Bhattacharya:2017chr}.
Since photon spheres allow naked singularities to mimic the optical
appearance of black holes, extensive research has focused on investigating
the gravitational lensing phenomena associated with naked singularities
\cite{Virbhadra:2002ju,Virbhadra:2007kw,Gyulchev:2008ff,Sahu:2012er,Shaikh:2019itn,Paul:2020ufc,Tsukamoto:2021fsz,Wang:2023jop,Chen:2023trn}.
Specifically, a photon emanating from the singularity might necessitate
an infinite coordinate time to reach a distant observer, akin to photons
departing from a black hole's event horizon \cite{Shaikh:2018lcc}.
This discovery, coupled with the presence of photon spheres, leads
to the absence of images of distant sources within the critical curve,
casting a shadow in naked singularity images. Nonetheless, in certain
spacetimes with naked singularities, photons can reach and escape
the singularity in a finite coordinate time \cite{Shaikh:2018lcc}.
In such scenarios, images of naked singularities captured by distant
observers hinge on the singularity's nature, yielding distinctions
from black hole images.

By minimally extending the Einstein field equations to include a massless
scalar field, the Janis-Newman-Winicour (JNW) solution naturally emerges
as the sole spherically symmetric solution permitted by the theory
\cite{Janis:1968zz}.  It encompasses the Schwarzschild black hole
as a specific limit, excluding the existence of other spherically
symmetric spacetimes featuring a regular event horizon. While the
literature contains several studies on the optical characteristics
of the JNW spacetime, including analyses of gravitational lensing,
relativistic images, accretion and shadows \cite{Virbhadra:1998dy,Virbhadra:2002ju,Virbhadra:2007kw,Gyulchev:2008ff,Chowdhury:2011aa,Gyulchev:2019tvk,Sau:2020xau,Martinez:2020hjm,Solanki:2021mkt,Chauvineau:2022bzg,Azizallahi:2023rrv},
it has seldom been explored whether photons can reach and depart from
the singularity in a finite coordinate time. If this occurs, how does
the singularity's nature influence the observational signatures of
JNW naked singularities?

To tackle these questions, we investigate the behavior of null geodesics
around the singularity and the gravitational lensing effects on distant
sources for the JNW naked singularity spacetime in this paper. The
subsequent sections of this paper are organized as follows: Section
\ref{sec:JNWS} offers a concise overview of the JNW solution and
examines the behavior of photon trajectories near the singularity.
In Section \ref{sec:LI}, we proceed with numerically simulating images
of a distant light source and a luminous celestial sphere. Following
that, Section \ref{sec:PS} analyzes images produced by a point-like
source. Lastly, our conclusions are presented in Section \ref{sec:Con}.
Throughout the paper, we adopt the convention $G=c=1$.

\section{Janis-Newman-Winicour Spacetime}

\label{sec:JNWS}

In this section, we provide a succinct overview of the JNW spacetime
and investigate the behavior of null geodesics within it, particularly
in proximity to the singularity. Subsequently, we explore light rays
in regularized singularity models, wherein the singularity is substituted
with a non-zero-sized regular core.

\subsection{Metric}

The JNW metric presents a static solution in Einstein-massless-scalar-field
models, characterized by equations of motion, 
\begin{align}
R_{\mu\nu} & -\frac{1}{2}Rg_{\mu\nu}=8\pi T_{\mu\nu},\nonumber \\
T_{\mu\nu} & =\nabla_{\mu}\Phi\nabla_{\nu}\Phi-\frac{1}{2}g_{\mu\nu}g^{\rho\sigma}\nabla_{\rho}\Phi\nabla_{\sigma}\Phi,\\
 & \nabla_{\mu}\nabla^{\mu}\Phi=0,\nonumber 
\end{align}
where $\Phi$ stands for the massless scalar field, and $T_{\mu\nu}$
is the corresponding energy-momentum tensor. The metric was initially
derived by Janis, Newman and Winicour, and is given as \cite{Janis:1968zz,Virbhadra:1995iy}
\begin{equation}
ds^{2}=-A\left(r\right)dt^{2}+B\left(r\right)dr^{2}+C\left(r\right)\left(d\theta^{2}+\sin^{2}\theta d\varphi^{2}\right),
\end{equation}
where the metric functions are
\begin{equation}
A\left(r\right)=B^{-1}\left(r\right)=\left(1-\frac{r_{g}}{r}\right)^{\gamma}\text{ and }C\left(r\right)=\left(1-\frac{r_{g}}{r}\right)^{1-\gamma}r^{2}.
\end{equation}
In addition, the scalar field is expressed as
\begin{equation}
\Phi=\frac{q}{r_{g}}\ln\left(1-\frac{r_{g}}{r}\right),
\end{equation}
where $q$ is the scalar charge.

\begin{figure}[ptb]
\includegraphics[scale=0.7]{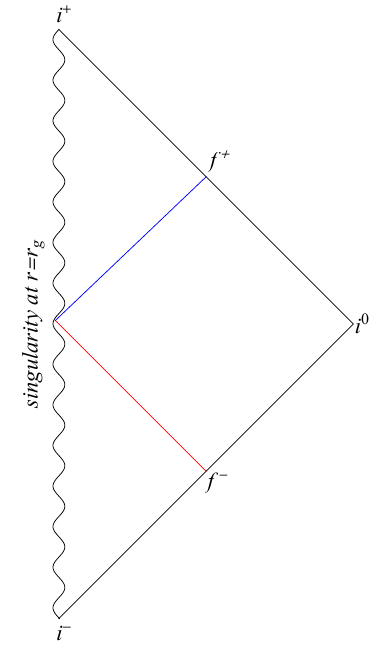}\caption{Penrose diagram for JNW naked singularity. The red and blue lines
denote light rays travelling toward and away from the singularity,
respectively.}
\label{fig:PDJNW}
\end{figure}

The JNW metric is characterized by two parameters, $\gamma$ and $r_{g}$,
which are related to the ADM mass $M$ and the scalar charge $q$
by \cite{Janis:1968zz}
\begin{equation}
\gamma=\frac{2M}{r_{g}}\text{ and }r_{g}=2\sqrt{M^{2}+q^{2}}.
\end{equation}
When $\gamma=1$, the JNW metric describes Schwarzschild black holes
with no scalar charge. For $0.5<\gamma<1$, the JNW metric represents
weakly naked singularity solutions with a non-trivial scalar field
profile. In this case, a naked curvature singularity occurs at $r=r_{g}$,
and a photon sphere is present at $r_{ps}=r_{g}\left(1+2\gamma\right)/2$.
However, the photon sphere disappears in the JNW metric when $0\leq\gamma<0.5$,
leading to distinct light propagations. Since a spacetime with photon
spheres can mimic black hole observations, such as black hole shadows,
the focus in this paper is on the JNW metric with $0.5<\gamma<1$.
The causal structure of the JNW naked singularity is depicted in the
corresponding Penrose diagram shown in FIG. \ref{fig:PDJNW}. The
timelike curvature singularity is located at $r=r_{g}$.

\subsection{Null Geodesics}

For a photon with 4-momentum vector $p^{\mu}=(\dot{t},\dot{r},\dot{\theta},\dot{\varphi})$,
where dots stand for derivative with respect to some affine parameter
$\lambda$, the null geodesic equations exhibit separability and can
be fully characterized by three conserved quantities, 
\begin{equation}
E=p_{\mu}\partial_{t}^{\mu}=-p_{t},\quad L_{z}=p_{\mu}\partial_{\varphi}^{\mu}=p_{\varphi},\quad L^{2}=K^{\mu\nu}p_{\mu}p_{\nu}=p_{\theta}^{2}+L_{z}^{2}\csc^{2}\theta,
\end{equation}
where the tensor $K^{\mu\nu}$ is a symmetric Killing tensor 
\begin{equation}
K=C^{2}\left(r\right)\left(d\theta\otimes d\theta+\sin^{2}\theta d\varphi\otimes d\varphi\right).
\end{equation}
Here, $E$, $L_{z}$ and $L$ denote the total energy, the angular
momentum parallel to the axis of symmetry, and the total angular momentum,
respectively. Consequently, the 4-momentum $p=p_{\mu}dx^{\mu}$ can
be expressed in terms of $E$, $L_{z}$ and $L$ as follows
\begin{align}
p & =-Edt\pm_{r}\sqrt{\mathcal{R}(r)}dr\pm_{\theta}\sqrt{\Theta(\theta)}d\theta+Ld\varphi,\nonumber \\
\mathcal{R}(r) & =B(r)\left[\frac{E^{2}}{A(r)}-\frac{L^{2}}{C(r)}\right]\text{ and }\Theta(\theta)=L^{2}-L_{z}^{2}\csc^{2}\theta,
\end{align}
where the choices of sign $\pm_{r}$ and $\pm_{\theta}$ depend on
the radial and polar directions of travel, respectively.

The null geodesic equations can then be written as
\begin{equation}
\dot{t}=\frac{E}{A(r)},\text{ }\dot{r}=\pm_{r}L\sqrt{b^{-2}-V_{\text{eff}}(r)},\text{ }\dot{\theta}=\pm_{\theta}\frac{\sqrt{L^{2}-L_{z}^{2}\csc^{2}\theta}}{C(r)},\text{ }\dot{\varphi}=\frac{L_{z}}{C(r)\sin^{2}\theta}\text{,}\label{eq:geo-eq}
\end{equation}
where $b\equiv L/E$ is the impact parameter, and the effective potential
of photons is defined as 
\begin{equation}
V_{\text{eff}}(r)=\frac{A\left(r\right)}{C(r)}.
\end{equation}
Particularly, the rate of change of radius $r$ with respect to time
$t$ can be expressed as 
\begin{equation}
\frac{dr}{dt}=\pm A(r)\sqrt{1-b^{2}V_{\text{eff}}(r)}.\label{eq:dr/dt}
\end{equation}
The plus and minus signs in the equation represent motion radially
outward and inward towards the naked singularity, respectively. It
is crucial to note that null geodesics are only permitted when $b^{-2}\geq V_{\text{eff}}(r)$.
The maximum of $V_{\text{eff}}(r)$ determines the locations of circular
light rays, which constitute the photon sphere. Interestingly, when
the impact parameter $b$ is smaller than the critical parameter $b_{c}$
(representing the impact parameter for photon trajectories on the
photon sphere), photons can cross the photon sphere and travel toward
the singularity.

To investigate whether a photon can reach and leave the singularity
at $r=r_{g}$ within a finite coordinate time (i.e., the proper time
measured by a distant observer), we first consider a radial light
ray with $b=0$, arriving at or departing from the singularity at
$t=0$. For this light ray, eqn. $\left(\ref{eq:dr/dt}\right)$ yields
\begin{equation}
t\left(r\right)=\pm\frac{r_{g}^{\gamma}\left(r-r_{g}\right)^{1-\gamma}}{\gamma-1}\text{ }{}_{2}F_{1}\left(1-\gamma,-\gamma,2-\gamma,1-\frac{r}{r_{g}}\right),\label{eq:trR}
\end{equation}
where plus and minus signs correspond to the outgoing and ingoing
cases, respectively. Here, $_{2}F_{1}\left(a,b,c;x\right)$ denotes
the hypergeometric function. The implication of eqn. $\left(\ref{eq:trR}\right)$
is twofold: firstly, photons emitted from distant sources reach the
singularity in a finite coordinate time, and secondly, photons leaving
the singularity reach distant observers in a finite coordinate time.
It is worth mentioning that $t\left(r\right)$ approaches infinity
in the limit of $\gamma=1$, which is consistent with expectations
for Schwarzschild black holes. For general null geodesics, eqn. $\left(\ref{eq:dr/dt}\right)$
describes the behavior of photons around the singularity at $r=r_{g}$,
yielding
\begin{equation}
\frac{dt}{dr}=\pm r_{g}^{\gamma}\left(r-r_{g}\right){}^{-\gamma}+\mathcal{O}\left(\left(r-r_{g}\right)^{1-\gamma}\right).\label{eq:dr/dt ep}
\end{equation}
The corresponding asymptotic solution is 
\begin{equation}
t\left(r\right)\sim\pm\frac{r_{g}^{\gamma}\left(r-r_{g}\right)^{1-\gamma}}{1-\gamma},
\end{equation}
which also shows that photons travel through the singularity in a
finite coordinate time.

Furthermore, in the vicinity of the singularity, the solutions of
the null geodesic equations $\left(\ref{eq:geo-eq}\right)$ take the
form
\begin{align}
r\left(\lambda\right) & =r_{g}\pm_{r}E\lambda+\mathcal{O}\left(\left\vert \lambda\right\vert ^{\frac{1}{2-2\gamma}}{}\right),\nonumber \\
t\left(\lambda\right) & =t_{0}\pm_{r}\frac{E^{1-\gamma}{}r_{g}^{\gamma}\left\vert \lambda\right\vert ^{1-\gamma}}{1-\gamma}+\mathcal{O}\left(\left\vert \lambda\right\vert {}^{1-\gamma}\right),\nonumber \\
\theta\left(\lambda\right) & =\theta_{0}\pm_{\theta}\sqrt{L^{2}-L_{z}^{2}\csc^{2}\theta_{0}}\frac{E^{\gamma-1}\left\vert \lambda\right\vert ^{\gamma}}{\gamma r_{g}^{1+\lambda}}+\mathcal{O}\left(\left\vert \lambda\right\vert {}^{\gamma}\right),\label{eq:geolamda}\\
\varphi\left(\lambda\right) & =\varphi_{0}+\frac{L_{z}E^{\gamma-1}\csc^{2}\theta_{0}\left\vert \lambda\right\vert ^{\gamma}}{\gamma r_{g}^{\gamma+1}}+\mathcal{O}\left(\left\vert \lambda\right\vert ^{\gamma}\right),\nonumber 
\end{align}
where $t_{0}$, $\theta_{0}$ and $\varphi_{0}$ are the integration
constants, and we assume $r\left(0\right)=r_{g}$. It shows the existence
of two classes of light rays: radially outgoing and ingoing light
rays, denoted as $+_{r}$ and $-_{r}$, respectively. To simplify,
we adopt $\lambda<0$ for ingoing light rays and $\lambda>0$ for
outgoing ones. As the affine parameter $\lambda$ approaches $0$
from the right and left, respectively, both outgoing and ingoing light
rays converge towards the singularity. In FIG. \ref{fig:PDJNW}, the
blue and red lines represent outgoing and ingoing light rays, respectively.

\subsection{Regularized Naked Singularity}

\label{sec:RNS}

\begin{figure}[ptb]
\includegraphics[scale=0.8]{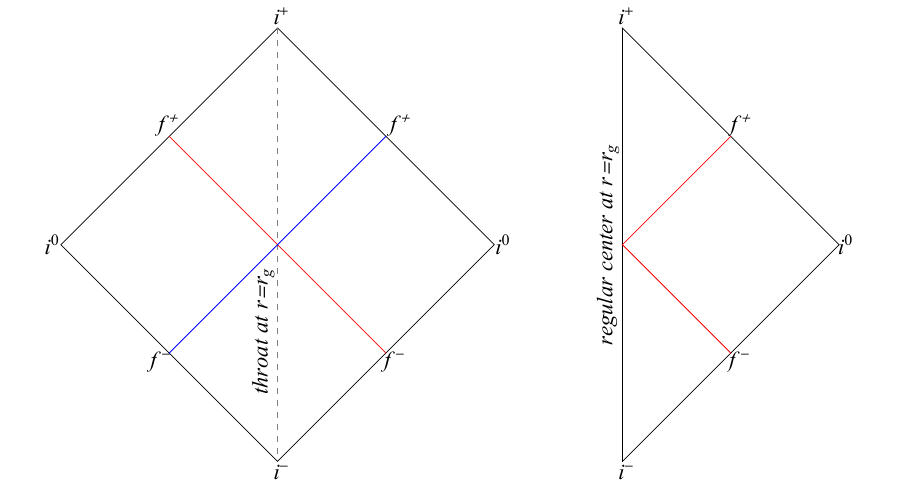}\caption{Penrose diagram for regularized JNW naked singularity with a wormhole
throat (\textbf{Left}) and a regular center (\textbf{Right}). }
\label{fig:PDRJNW}
\end{figure}

As mentioned earlier, photons with a sufficiently small impact parameter
can reach the singularity in a finite coordinate time, prompting inquiry
into their fate at the singularity as observed by distant observers.
However, the existence of the singularity indicates a breakdown in
the applicability of general relativity around it, which has ignited
the pursuit of a theory of quantum gravity. Given the lack of a definitive
quantum gravity framework, researchers frequently resort to effective
models to regularize such singularities, thereby facilitating investigations
into the characteristics of null geodesics at these points. These
effective models often employ a regularized singularity metric that
can describe spacetime like wormholes or non-singular manifolds. An
example in the former category is the modified JNW geometry through
the application of the Simpson and Visser (SV) method \cite{Pal:2022cxb}.
Specifically, the SV-modified JNW geometry is derived by substituting
$r$ with $\sqrt{r^{2}+\epsilon^{2}}$, where $\epsilon$ represents
the SV parameter. When two such SV-modified JNW metrics are connected
by a thin shell at $r=r_{g}$, the resulting spacetime forms a traversable
wormhole with a throat located at $r=r_{g}$. The corresponding Penrose
diagram is depicted in the left panel of FIG. \ref{fig:PDRJNW}, where
the red and blue lines illustrate that light rays moving toward $r=r_{g}$
will traverse the throat and enter another universe.

Of particular interest, the regularization of singularities can be
achieved by incorporating higher-order curvature terms, such as the
complete $\alpha^{\prime}$ corrections of string theory \cite{Wang:2019kez,Wang:2019dcj,Ying:2021xse}.
This approach leads to a regularized singularity spacetime that maintains
regularity throughout, and $r=r_{g}$ serves as the center of this
well-behaved manifold. Moreover, apart from a vicinity around the
center, the singularity spacetime closely approximates the regularized
singularity spacetime. Therefore, photons with sufficiently small
impact parameters are able to traverse the neighborhood of the center
and reach distant observers. The right panel of FIG. \ref{fig:PDRJNW}
shows the corresponding Penrose diagram, where a red line illustrates
a light ray passing through the neighborhood of the center. Note that
this light ray comprises both radially outgoing and ingoing branches.
Typically, a specific formulation of the regularized singularity spacetime
is necessary to establish the connection between the ingoing and outgoing
branches.

For the sake of simplicity, we construct a regularized singularity
spacetime by matching the JNW metric with a regular spacetime, introducing
a thin shell centered at $r=r_{g}$ with a tiny yet non-zero radius
$\epsilon$. The metric of the regular spacetime within this thin
shell is described by 
\begin{equation}
ds_{\text{in}}^{2}=-A_{\text{in}}\left(r\right)dt^{2}+B_{\text{in}}\left(r\right)dr^{2}+C_{\text{in}}\left(r\right)\left(d\theta^{2}+\sin^{2}\theta d\varphi^{2}\right).
\end{equation}
In the vicinity of $r=r_{g}$, the metric functions are expanded in
the following manner,
\begin{align}
A_{\text{in}}\left(r\right) & =a_{0}+a_{1}\left(r-r_{g}\right)+\cdots,\nonumber \\
B_{\text{in}}\left(r\right) & =b_{0}+b_{1}\left(r-r_{g}\right)+\cdots,\\
C_{\text{in}}\left(r\right) & =\left(r-r_{g}\right)^{2}\left[c_{0}+c_{1}\left(r-r_{g}\right)+\cdots\right],\nonumber 
\end{align}
where the equality of $b_{0}$ and $c_{0}$ arises from the absence
of a conical singularity at $r=r_{g}$. For a light ray entering the
regular core from the JNW spacetime, the corresponding energy $E_{\text{in}}$
and angular momentum $L_{\text{in}}$ in the regular core are related
to the energy $E$ and angular momentum $L$ in the JNW spacetime
as \cite{Wang:2020emr,Chen:2022tog},
\begin{equation}
E_{\text{in}}=\sqrt{\frac{A_{\text{in}}\left(r_{g}+\epsilon\right)}{A\left(r_{g}+\epsilon\right)}}E\text{ and }L_{\text{in}}=\sqrt{\frac{C_{\text{in}}\left(r_{g}+\epsilon\right)}{C\left(r_{g}+\epsilon\right)}}L\text{.}
\end{equation}

Due to the spherical symmetry, we restrict the light ray to the equatorial
plane and assume its entry and exit from the thin shell occur at $\varphi=\varphi_{1}$
and $\varphi_{2}$, respectively. Beyond the thin shell, in the JNW
spacetime, the light ray follows the trajectory given by eqn. $\left(\ref{eq:geolamda}\right)$,
encompassing both ingoing and outgoing branches that terminates at
$\varphi=\varphi_{1}$ and departs from $\varphi=\varphi_{2}$ on
the thin shell, respectively. The deflection angle $\Delta\varphi=\varphi_{2}-\varphi_{1}$,
due to the influence of the regular core, is quantified as
\begin{equation}
\Delta\varphi=2\int_{b_{\text{in}}}^{r_{g}+\epsilon}\frac{\sqrt{A_{\text{in}}\left(r\right)B_{\text{in}}\left(r\right)}dr}{C_{\text{in}}(r)\sqrt{b_{\text{in}}^{-2}-A_{\text{in}}\left(r\right)/C_{\text{in}}\left(r\right)}},
\end{equation}
where $b_{\text{in}}=L_{\text{in}}/E_{\text{in}}$. As $\epsilon$
approaches zero, 
\begin{equation}
\Delta\varphi\rightarrow2\arccos\left(\epsilon^{\gamma-1/2}r_{g}^{-\gamma-1/2}b\right)\rightarrow\pi,
\end{equation}
illustrating that the entry and exit points on the thin shell are
antipodal opposite. Furthermore, when the singularity is modeled with
an infinitesimally small regular core, the portion of the light ray
within the thin shell can be safely disregarded. Thus, the light ray
is well approximated by the combination of the ingoing and outgoing
branches in the JNW metric outside the thin shell. Since these branches
intersect with the thin shell at exactly two antipodal points, their
connection is given by 
\begin{equation}
\theta\rightarrow\pi-\theta\text{ and }\varphi\rightarrow\varphi+\pi.\label{eq:cc}
\end{equation}
In short, the condition $\left(\ref{eq:cc}\right)$ and the conservation
of $E$, $L_{z}$ and $L$ determine the corresponding outgoing branch
for a given ingoing branch.

Given its stronger theoretical motivation and its potential for generating
a richer set of observational outcomes, we focus on the case of a
regularized JNW singularity with a regular center throughout the remainder
of this paper. Moreover, the observations related to the wormhole
case can be encompassed within the context of the regular center case,
an aspect that we will briefly address in the concluding section.

\section{Lensing Images}

\label{sec:LI}

This section presents a numerical simulation of gravitational lensing
images produced by a distant luminous object and a celestial sphere
around JNW naked singularities, aiming to illustrate the gravitational
lensing phenomenon. The observer and the center of the luminous object
are positioned on the equatorial plane at $\left(r,\phi\right)=\left(10M,\pi\right)$
and $\left(25M,\pi/6\right)$, respectively, while the luminous celestial
sphere is placed at $r=25M$. Here, $M$ represents the mass of the
JNW metric. To generate observational images, we vary the observer's
viewing angle and numerically integrate $2000\times2000$ photon trajectories
until they intersect with the celestial sphere. For a detailed explanation
of the numerical implementation, interested readers can refer to \cite{Chen:2023trn}.

The strong deflection around the photon sphere causes light rays to
produce multiple images of the distant object when they orbit the
photon sphere multiple times. We conveniently classify these images
based on the number of orbits, denoted by $n$. Moreover, the positive
and negative signs of $n$ indicate whether the orbiting occurs in
the counter-clockwise or clockwise directions. As anticipated, light
rays can circle around the photon sphere slightly outside of it, generating
images similar to the Schwarzschild black hole case. Interestingly,
our earlier discussion revealed that light rays, being able to traverse
the singularities, can also orbit the photon sphere from inside, resulting
in a new set of images. Throughout the remainder of this paper, we
use the notations $n^{>}$ and $n^{<}$ to refer to the $\left\vert n\right\vert ^{\text{th}}$-order
images produced by light rays orbiting $n$ times outside and inside
the photon sphere, respectively.

\begin{figure}[ptb]
\includegraphics[scale=0.7]{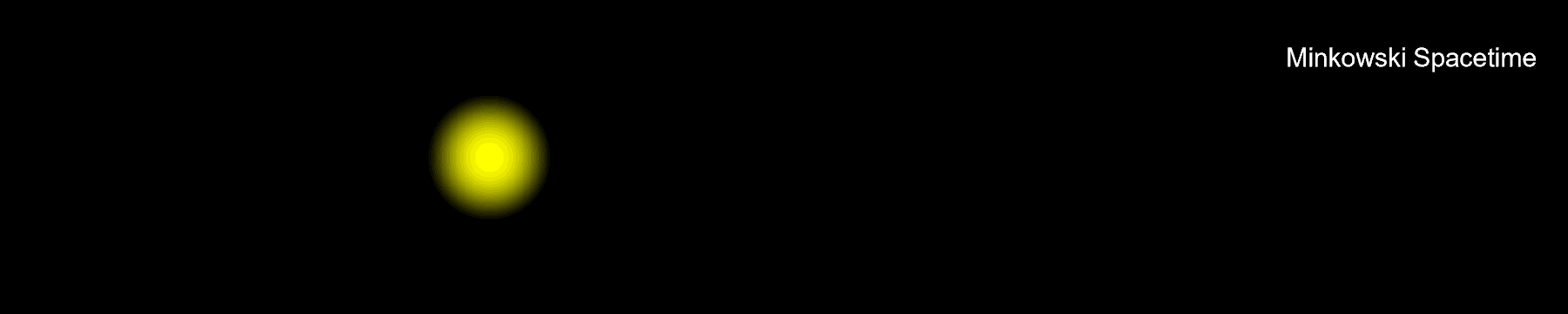} \includegraphics[scale=0.7]{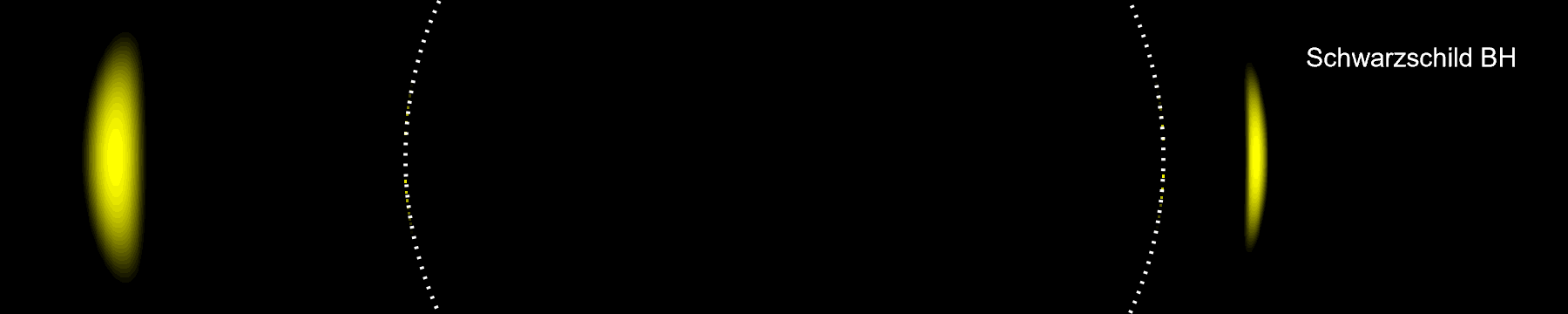}
\includegraphics[scale=0.7]{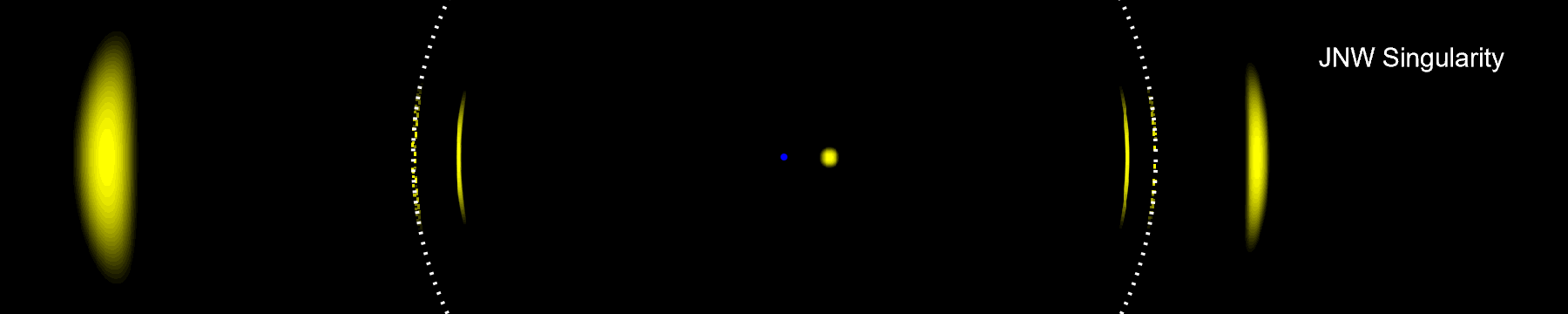} \includegraphics[scale=0.7]{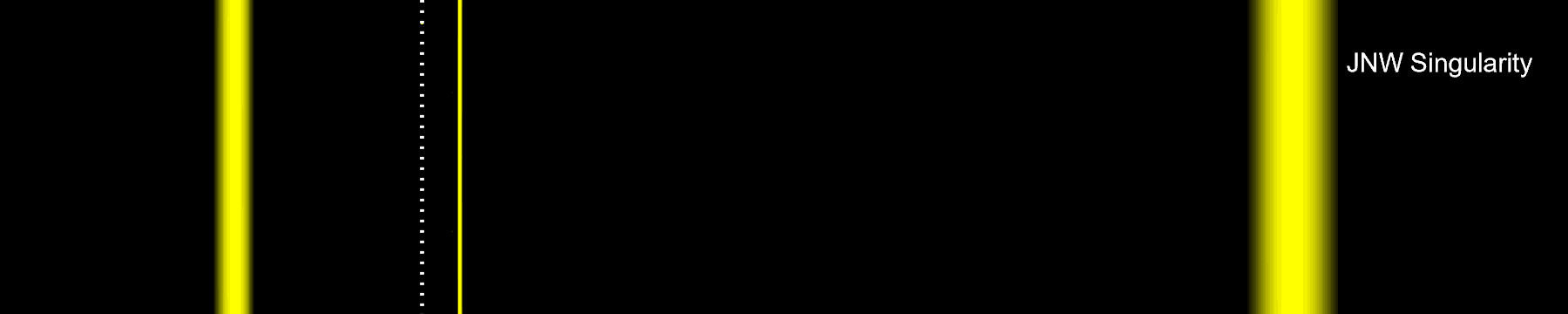}\caption{Images of a luminous object centered at $x_{\text{s}}^{\mu}=(0,25M,\pi/2,\pi/6)$
as viewed by an observer positioned at $x_{\text{o}}^{\mu}=(0,10M,\pi/2,\pi)$
in Minkowski spacetime, a Schwarzschild black hole, and a JNW singularity
with $\gamma=2/3$. The critical curve, formed by light rays escaping
from the photon sphere, is depicted with dashed white lines, while
the JNW singularity is marked by a blue dot. Unlike the Schwarzschild
black hole, the JNW singularity displays additional images of the
object inside the critical curve, due to the capability of light rays
to pass through the singularity. A zoom-in view near the JNW singularity's
critical curve is presented in the bottom panel, showing one higher-order
image outside the critical curve and two higher-order images inside
it.}
\label{fig:HS}
\end{figure}

\begin{figure}[ptb]
\includegraphics[scale=1.1]{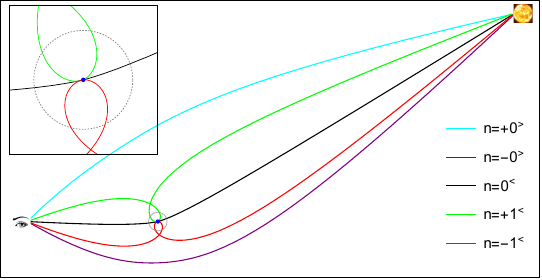} \includegraphics[scale=1.1]{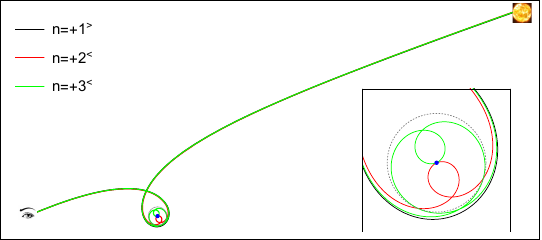}
\caption{\textbf{Upper}: Light rays responsible for generating the object images
in the third panel (located just above the bottom panel) of FIG. \ref{fig:HS},
where the images with $n=+0^{>}$, $+1^{<}$, $0^{<}$, $-1^{<}$
and $-0^{>}$ are presented from left to right. \textbf{Lower}: Light
rays that produce the object images in the bottom panel of FIG. \ref{fig:HS},
where the images with $n=+1^{>}$, $+3^{<\text{ }}$and $+2^{<\text{ }}$
are displayed from left to right. In $n$, the number denotes the
number of orbits of light rays, while the $+$ and $-$ signs indicate
the counter-clockwise and clockwise directions, respectively. Additionally,
the $>$ and $<$ symbols correspond to orbiting outside and inside
the photon sphere, respectively. The dashed circular lines represent
the photon sphere.}
\label{fig:LT}
\end{figure}

FIG. \ref{fig:HS} displays images of the luminous object in three
different scenarios: Minkowski spacetime, a Schwarzschild black hole,
and a JNW naked singularity with $\gamma=2/3$. In the figure, the
dashed white lines represent the critical curve, which is formed by
photons originating from the photon sphere. The images produced by
light rays with $n^{>}$ and $n^{<}$ lie outside and inside the critical
curve, respectively. Outside the critical curve, both the Schwarzschild
black hole and the JNW singularity cases exhibit two observable images:
$n=+0^{>}$ and $-0^{>}$, referred to as the primary and secondary
images, respectively. These images can be analyzed using the weak
lensing approximation. However, for the Schwarzschild black hole,
no images are visible inside the critical curve due to the presence
of the event horizon. In contrast, the JNW naked singularity presents
an image with $n=0^{<}$ near the center and two additional images
with $n=\pm1^{<}$ within the critical curve. The light rays responsible
for generating these images are illustrated in the upper panel of
FIG. \ref{fig:LT}. It should be noted that the higher-order images
are located so close to the critical curve that they cannot be resolved
in the two middle panels of FIG. \ref{fig:HS}. To provide a closer
view of the critical curve region, the bottom panel of FIG. \ref{fig:HS}
zooms in and displays the $n=+1^{>}$, $+3^{<\text{ }}$and $+2^{<\text{ }}$
images from left to right. Furthermore, the light rays responsible
for producing these three images are depicted in the lower panel of
FIG. \ref{fig:LT}. It is worth emphasizing that, except for the $n=0^{<}$
image, all other images are significantly distorted due to the effects
of gravitational lensing.

\begin{figure}[ptb]
\includegraphics[scale=0.076]{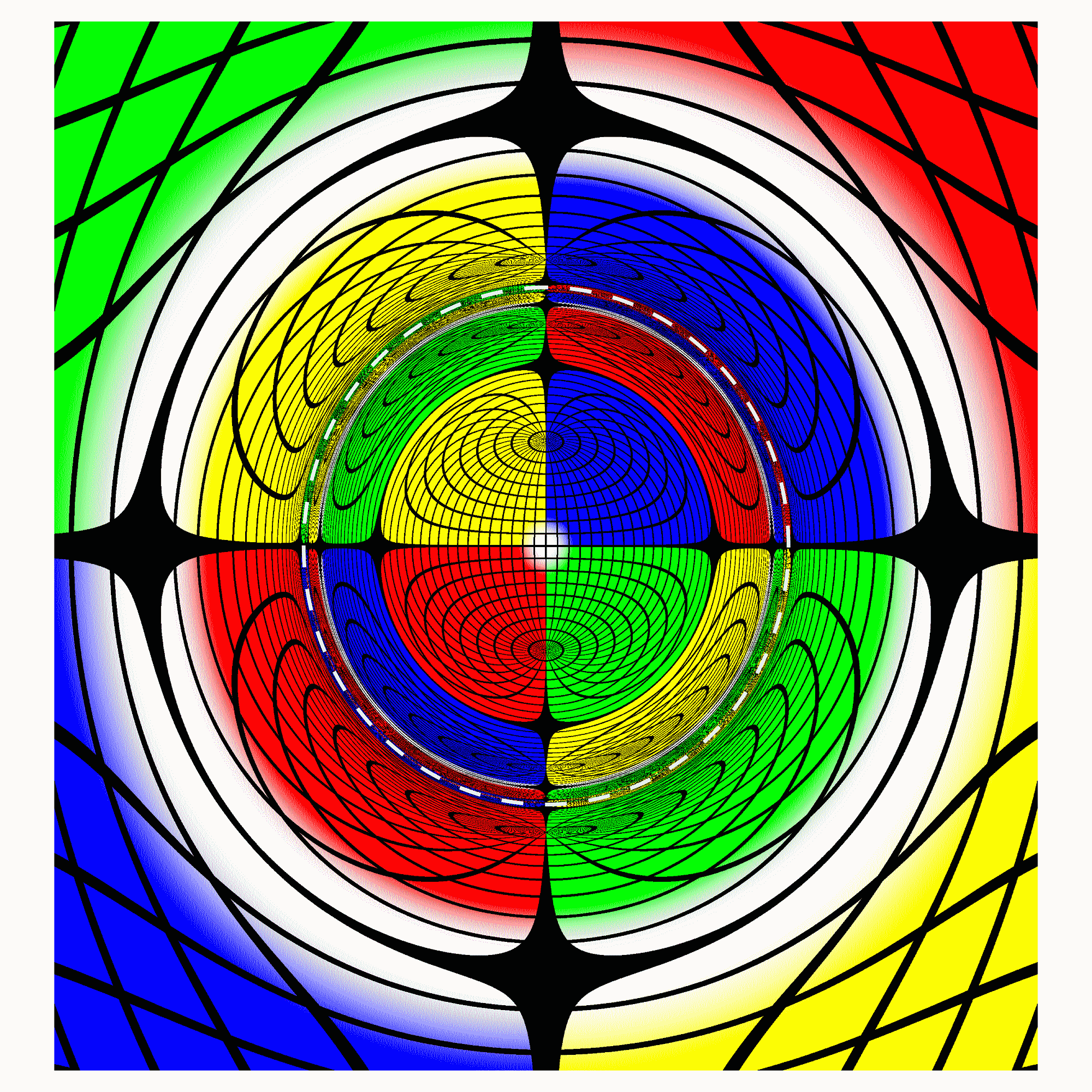} \includegraphics[scale=0.076]{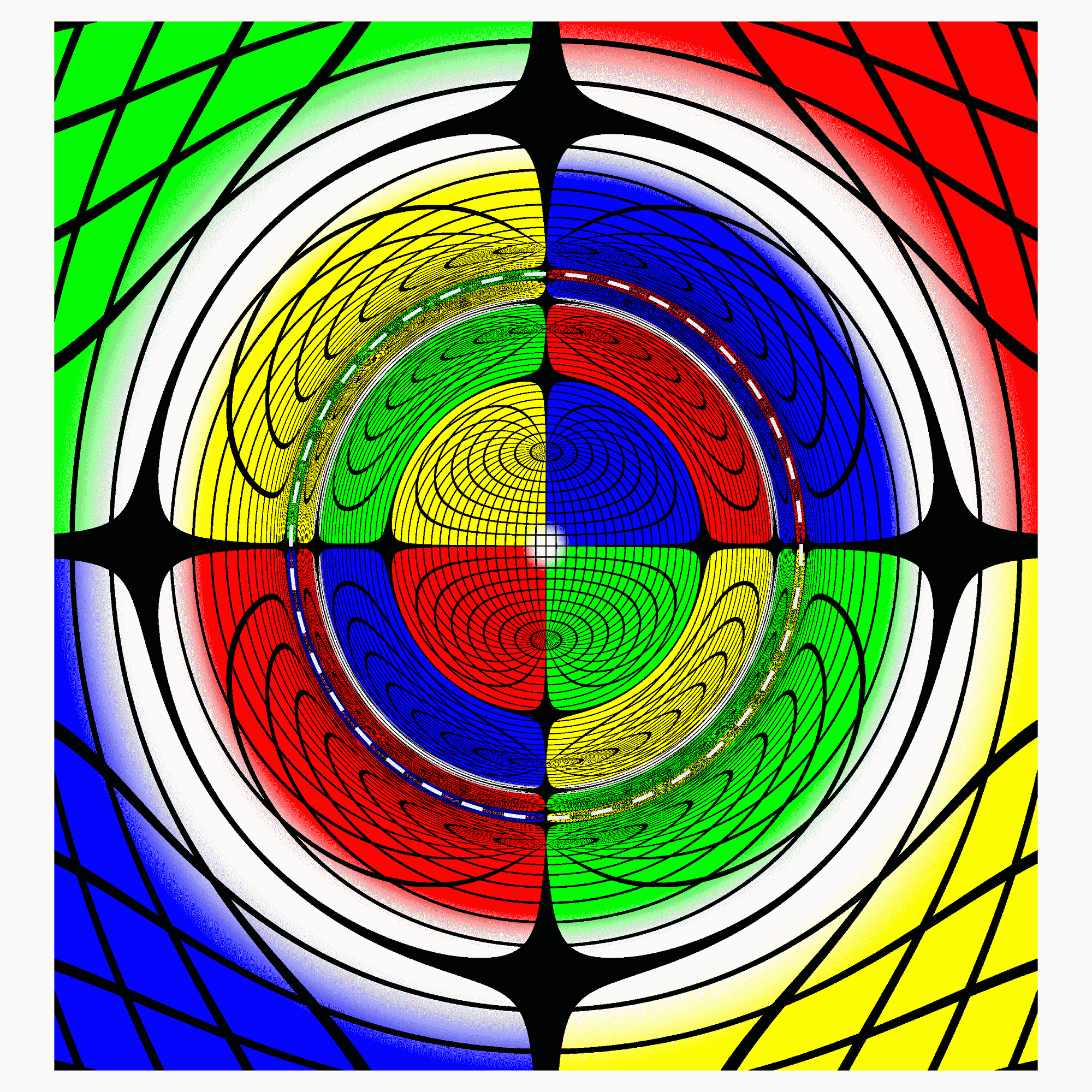}
\includegraphics[scale=0.076]{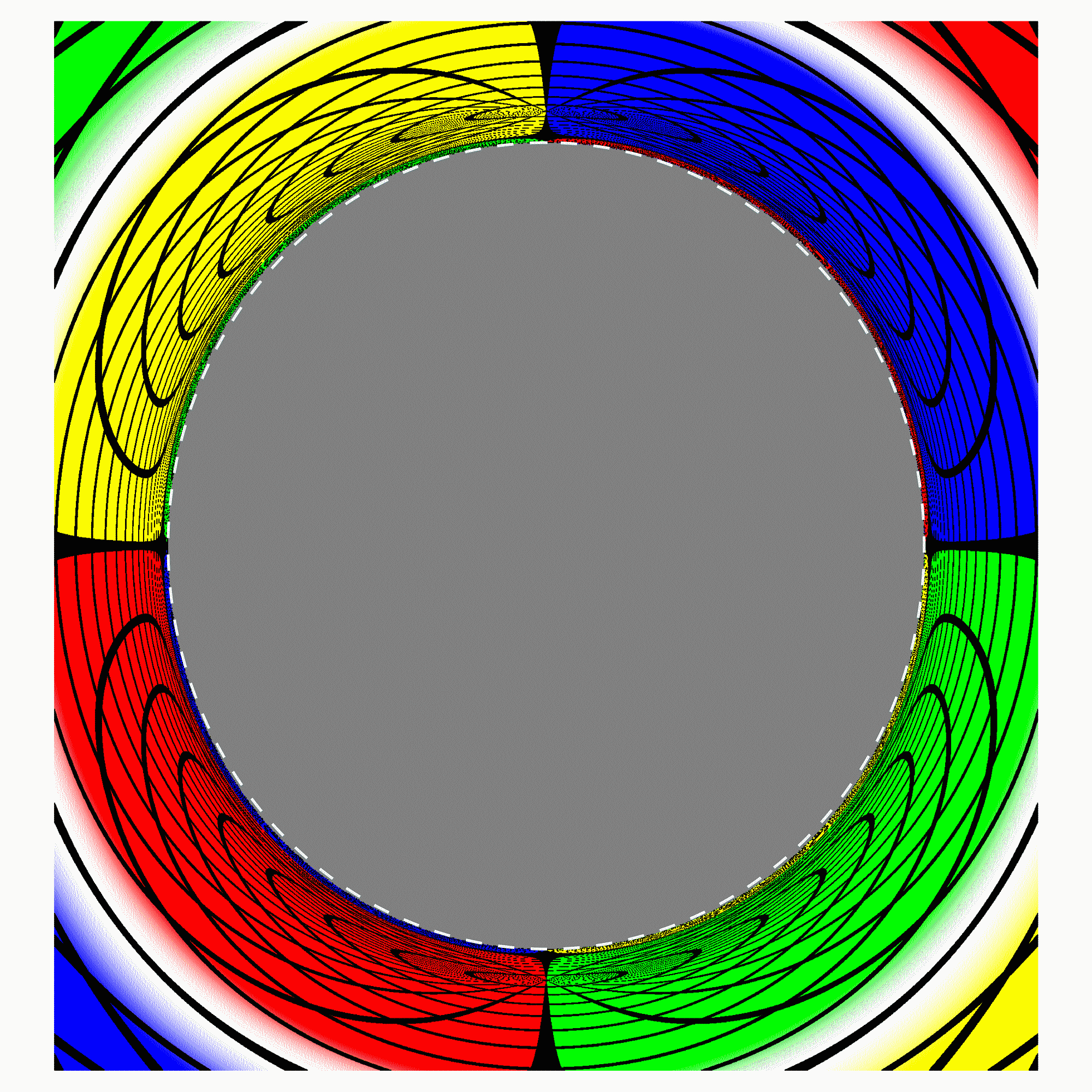}\caption{Images of a celestial sphere located at $r=25M$ in the JNW metric
with $\gamma=0.6$ (\textbf{Left}), $0.9$ (\textbf{Middle}) and $1$
(\textbf{Right}). The observer's location is $x_{\text{o}}^{\mu}=(0,10M,\pi/2,\pi)$,
and the field of view spans $2\pi/5$. The dashed lines represent
the critical curve. In the case of $\gamma=1$, the JNW metric corresponds
to a Schwarzschild black hole, and thus, the right panel includes
the black hole shadow depicted as a gray area. For $\gamma=0.6$ and
$0.9$, the image inside the critical curve is formed by light rays
passing through the singularity. A central white dot is visible, surrounded
by two white Einstein rings.}
\label{fig:CS}
\end{figure}

Furthermore, FIG. \ref{fig:CS} illustrates images of the celestial
sphere in both a Schwarzschild black hole and JNW singularities. The
celestial sphere is divided into four quadrants, each distinguished
by a different color, and a white dot is placed in front of the observer.
Moreover, a grid of black lines, with adjacent lines separated by
$\pi/18$, is overlaid to represent constant longitude and latitude.
In the images shown in FIG. \ref{fig:CS}, the dashed circular lines
represent the critical curve. Outside the critical curve, the celestial
sphere images in JNW singularities bear a resemblance to those observed
in the Schwarzschild black hole spacetime. However, while shadows
are observed in the black hole image, the celestial sphere images
persist within the critical curve for JNW singularities. This unique
feature is attributed to the transparency of the singularity, allowing
light to traverse through it.

Outside the critical curve, we observe a white Einstein ring generated
by the white dot placed on the celestial sphere in both the Schwarzschild
black hole and JNW singularity cases. However, in JNW singularities,
an additional Einstein ring appears within the critical curve, representing
photons that pass through the singularity with angular coordinate
changes of $\Delta\varphi=3\pi$. Furthermore, light rays passing
through the singularity and undergoing an angular coordinate change
of $\Delta\varphi=\pi$ result in a white dot positioned at the center
of the image.

\section{Point-like Source}

\label{sec:PS}

In order to investigate point-source gravitational lensing, we adopt
an idealized thin lens model that assumes a high degree of alignment
among the source, lens and observer. The lens equation, as presented
in \cite{mollerach2002gravitational,Virbhadra:1999nm}, is expressed
as 
\begin{equation}
\beta=\vartheta-\frac{D_{LS}}{D_{OS}}\Delta\alpha,\label{eq:beta}
\end{equation}
where $\beta$ represents the angular separation between the source
and the lens, $\vartheta$ denotes the angular separation between
the lens and the image, and $\Delta\alpha$ represents the offset
of the total deflection angle $\alpha$ after accounting for all the
windings experienced by the photon. Here, the distances $D_{OL}$,
$D_{LS}$ and $D_{OS}$ correspond to the observer-lens, lens-source
and observer-source distances, respectively. Furthermore, we define
the magnification $\mu$ of an image as the ratio of the image's flux
to the flux of the unlensed source. This ratio is determined by the
solid angles of the image and the unlensed source measured by the
observer, resulting in the following expression \cite{mollerach2002gravitational,Virbhadra:2022iiy}
\begin{equation}
\mu=\mu^{r}\mu^{t}=\frac{\vartheta}{\beta}\frac{d\vartheta}{d\beta}.\label{eq:mu}
\end{equation}
In the above equation, the factors $\mu^{r}\equiv$ $d\vartheta/d\beta$
and $\mu^{t}\equiv$ $\vartheta/\beta$ represent the radial and tangential
magnifications of the image, respectively. Additionally, the sign
of the magnification factors determines the parity of the image.

By taking advantage of the spherical symmetry, our computations of
the deflection angle $\alpha$ are confined to the equatorial plane.
Within the thin-lens approximation, the deflection angle $\alpha$
is governed by the expression from \cite{Virbhadra:1999nm}, 
\begin{equation}
\alpha=I(b)-\pi,
\end{equation}
where $I(b)$ represents the change in $\varphi$, and $b$ denotes
the impact parameter related to $\vartheta$ through the equation
$b=D_{OL}\vartheta$. When a photon approaches a turning point at
$r=r_{0}$ and then gets deflected towards a distant observer, the
integral $I(b)$ is given by 
\begin{equation}
I(b)=2\int_{r_{0}}^{\infty}\frac{dr}{C(r)\sqrt{b^{-2}-V_{\text{eff}}\left(r\right)}},\label{eq:IR}
\end{equation}
where $r_{0}$ is determined by $V_{\text{eff}}\left(r_{0}\right)=b^{-2}$.
On the other hand, if the photon passes through the singularity at
$r=r_{g}$, the azimuthal angle $\varphi$ increases by $\pi$, leading
to the expression, 
\begin{equation}
I(b)=2\int_{r_{g}}^{\infty}\frac{dr}{C(r)\sqrt{b^{-2}-V_{\text{eff}}\left(r\right)}}+\pi.\label{eq:IT}
\end{equation}

\subsection{Images Outside the Critical Curve}

\begin{table}[ptb]
\begin{tabular}{ccccc}
\hline 
$\gamma$  & 0.6  & 0.7  & 0.8  & 0.9\tabularnewline
\hline 
\hline 
$\vartheta_{\pm\infty}$  & $\pm25.1927$  & $\pm26.5221$  & $\pm27.2768$  & $\pm27.7592$\tabularnewline
\hline 
$\vartheta_{+0^{>}}$  & $5.4121\times10^{6}$  & $5.4121\times10^{6}$  & $5.4121\times10^{6}$  & $5.4121\times10^{6}$\tabularnewline
$\Delta\vartheta_{+1^{>}}$  & $3.9857\times10^{-2}$  & $3.6895\times10^{-2}$  & $3.5898\times10^{-2}$  & $3.5424\times10^{-2}$\tabularnewline
$\Delta\vartheta_{+2^{>}}$  & $7.4431\times10^{-5}$  & $6.8900\times10^{-5}$  & $6.7037\times10^{-5}$  & $6.6152\times10^{-5}$\tabularnewline
$\Delta\vartheta_{+3^{>}}$  & $1.3900\times10^{-7}$  & $1.2867\times10^{-7}$  & $1.2519\times10^{-7}$  & $1.2354\times10^{-7}$\tabularnewline
\hline 
$\vartheta_{-0^{>}}$  & $-4.1204\times10^{5}$  & $-4.1204\times10^{5}$  & $-4.1204\times10^{5}$  & $-4.1204\times10^{5}$\tabularnewline
$\Delta\vartheta_{-1^{>}}$  & $-3.9853\times10^{-2}$  & $-3.6892\times10^{-2}$  & $-3.5894\times10^{-2}$  & $-3.5420\times10^{-2}$\tabularnewline
$\Delta\vartheta_{-2^{>}}$  & $-7.4424\times10^{-5}$  & $-6.8893\times10^{-5}$  & $-6.7031\times10^{-5}$  & $-6.6146\times10^{-5}$\tabularnewline
$\Delta\vartheta_{-3^{>}}$  & $-1.3898\times10^{-7}$  & $-1.2865\times10^{-7}$  & $-1.2516\times10^{-7}$  & $-1.2352\times10^{-7}$\tabularnewline
\hline 
$\mu_{+0^{>}}$  & $1.0058$  & $1.0058$  & $1.0058$  & $1.0058$\tabularnewline
$\mu_{+1^{>}}$  & $1.9502\times10^{-18}$  & $1.9002\times10^{-18}$  & $1.9013\times10^{-18}$  & $1.9093\times10^{-18}$\tabularnewline
$\mu_{+2^{>}}$  & $3.6362\times10^{-21}$  & $3.5436\times10^{-21}$  & $3.5459\times10^{-21}$  & $3.5609\times10^{-21}$\tabularnewline
$\mu_{+3^{>}}$  & $6.7903\times10^{-24}$  & $6.6174\times10^{-24}$  & $6.6217\times10^{-24}$  & $6.6498\times10^{-24}$\tabularnewline
\hline 
$\mu_{-0^{>}}$  & $-5.8297\times10^{-3}$  & $-5.8297\times10^{-3}$  & $-5.8297\times10^{-3}$  & $-5.8297\times10^{-3}$\tabularnewline
$\mu_{-1^{>}}$  & $-1.9502\times10^{-18}$  & $-1.9002\times10^{-18}$  & $-1.9013\times10^{-18}$  & $-1.9093\times10^{-18}$\tabularnewline
$\mu_{-2^{>}}$  & $-3.6362\times10^{-21}$  & $-3.5436\times10^{-21}$  & $-3.5459\times10^{-21}$  & $-3.5609\times10^{-21}$\tabularnewline
$\mu_{-3^{>}}$  & $-6.7903\times10^{-24}$  & $-6.6174\times10^{-24}$  & $-6.6217\times10^{-24}$  & $-6.6498\times10^{-24}$\tabularnewline
\hline 
\end{tabular}\caption{The angular position $\vartheta_{\pm n^{>}}$ and the magnification
factor $\mu_{\pm n^{>}}$ of images outside the critical curve for
a point-like source in JNW singularities with varying $\gamma$. Here,
$\Delta\vartheta_{\pm n^{>}}\equiv\vartheta_{\pm n^{>}}-\vartheta_{\pm\infty}$
represents the angular separation between the $n^{\text{th}}$-order
relativistic images and the critical curve. The superscript $>$ indicates
that the images are produced by light rays with $b>b_{c}$ and therefore
reside outside the critical curve. The values $M=4.31\times10^{6}M_{\astrosun}$,
$D_{OL}=D_{LS}=7.86$ kpc and $\beta=5$ arcsec are used. All angles
are expressed in units of microarcseconds.}
\label{table:out}
\end{table}

As depicted above, the images located outside the critical curve are
produced by light rays reaching a turning point at $r=r_{0}$, which
lies outside the photon sphere. To calculate the angular position
$\vartheta_{\pm0^{>}}$ and magnification factor $\mu_{\pm0^{>}}$
for the primary image with $n=+0^{>}$ and the secondary one with
$n=-0^{>}$, we employ eqn. $\left(\ref{eq:IR}\right)$ to numerically
determine $I(b)$ and $dI\left(b\right)/db$. By utilizing the resulting
$I(b)$ and $dI\left(b\right)/db$, eqns. $\left(\ref{eq:beta}\right)$
and $\left(\ref{eq:mu}\right)$ provide the desired $\vartheta_{\pm0^{>}}$
and $\mu_{\pm0^{>}}$. Additionally, our findings reveal that $\mu_{\pm0^{>}}^{t}\sim\beta^{-1}$
and $\mu_{\pm0^{>}}^{r}\sim\mathcal{O}\left(1\right)$, as expected
from weak gravitational lensing. When $\beta\ll1$, the magnitude
of $\mu_{\pm0^{>}}^{t}$ greatly exceeds that of $\mu_{\pm0^{>}}^{r}$,
resulting in significantly distorted images.

In the case of relativistic images with $\left\vert n\right\vert \geq1$,
their impact parameter $b$ closely approaches the critical impact
parameter $b_{c}$, allowing us to expand $I(b)$ around $b=b_{c}$.
In this strong deflection limit, the total deflection angle $\alpha$
is expressed as \cite{Bozza:2002zj,Tsukamoto:2016jzh}
\begin{equation}
\alpha=-\bar{a}^{>}\ln\left(b/b_{c}-1\right)+\bar{b}^{>}+\mathcal{O}\left(\left(b/b_{c}-1\right)\ln\left(b/b_{c}-1\right)\right),\label{eq:alpha-a}
\end{equation}
where
\begin{equation}
\bar{a}^{>}=1\text{ and }\bar{b}^{>}=-\pi+I_{R}^{>}+\ln\left[\frac{2\left(2\gamma+1\right)}{2\gamma-1}\right].
\end{equation}
Here, the term $I_{R}^{>}$ represents a regular integral that can
be computed numerically. Using eqns. $\left(\ref{eq:beta}\right)$
and $\left(\ref{eq:mu}\right)$, we can solve for the angular position
$\vartheta_{\pm n^{>}}$ and magnification factor $\mu_{\pm n^{>}}$
of $n^{\text{th}}$-order relativistic images. Specifically, the angular
position $\vartheta_{\pm n^{>}}$ is given by \cite{Bozza:2002zj}
\begin{equation}
\vartheta_{\pm n^{>}}=\vartheta_{\pm n^{>}}^{0}+\frac{b_{c}e_{n}^{>}D_{OS}}{\bar{a}^{>}D_{LS}D_{OL}}\left(\beta-\vartheta_{\pm n^{>}}^{0}\right),\label{eq:theta-out}
\end{equation}
where $e_{n}^{>}=e^{\frac{\bar{b}^{>}-2\pi n}{\bar{a}^{>}}}$, and
$\vartheta_{\pm n^{>}}^{0}$, satisfying $\alpha\left(\vartheta_{\pm n^{>}}^{0}\right)=\pm2n\pi$,
is given by 
\begin{equation}
\vartheta_{\pm n^{>}}^{0}=\pm\frac{b_{c}}{D_{OL}}\left(1+e_{n}^{>}\right).\label{eq:Ering-a}
\end{equation}
The factors $\mu^{t}$ and $\mu^{r}$ are then expressed as
\begin{equation}
\mu_{\pm n^{>}}^{t}=\pm\frac{b_{c}\left(1+e_{n}^{>}\right)}{\beta D_{OL}}\text{ and }\mu_{\pm n^{>}}^{r}=\frac{b_{c}e_{n}^{>}D_{OS}}{\bar{a}^{>}D_{LS}D_{OL}},
\end{equation}
which gives the magnification factor $\mu_{\pm n^{>}}=\mu_{\pm n^{>}}^{r}\mu_{\pm n^{>}}^{t}$.
As $\left\vert \mu_{\pm n^{>}}^{t}\right\vert \gg\left\vert \mu_{\pm n^{>}}^{r}\right\vert $,
the relativistic images are significantly stretched along the critical
curve.

To numerically estimate $\vartheta_{\pm n^{>}}$ and $\mu_{\pm n^{>}}$
in an astrophysical setting, we model the supermassive black hole
Sgr A{*} located at the center of our Galaxy as a JNW singularity.
Specifically, we assume a mass of $M=4.31\times10^{6}M_{\astrosun}$
and a lens-source distance of $D_{OL}=7.86$ kpc. Additionally, a
source is positioned at $D_{LS}=7.86$ kpc with an angular separation
of $\beta=5$ arcsec. Table \ref{table:out} presents $\vartheta_{\pm0^{>}}$
and $\mu_{\pm0^{>}}$ for the primary and secondary images, along
with $\Delta\vartheta_{\pm n^{>}}\equiv\vartheta_{\pm n^{>}}-\vartheta_{\pm\infty}$
and $\mu_{\pm n^{>}}$ for the relativistic images, considering various
values of $\gamma$ in JNW singularities. Here, $\vartheta_{\pm\infty}=\lim\limits _{n\rightarrow\infty}\vartheta_{\pm n^{>}}^{0}=\pm b_{c}/D_{OL}$
represents the angular position of the critical curve. The results
indicate that the angular position $\vartheta_{\pm n^{>}}$ and magnification
factor $\mu_{\pm n^{>}}$ of the images outside the critical curve
show little sensitivity to $\gamma$.

\subsection{Images Inside the Critical Curve}

\begin{table}[ptb]
\begin{tabular}{ccccc}
\hline 
$\gamma$  & 0.6  & 0.7  & 0.8  & 0.9\tabularnewline
\hline 
\hline 
$\vartheta_{\pm\infty}$  & $\pm25.1927$  & $\pm26.5221$  & $\pm27.2768$  & $\pm27.7592$\tabularnewline
\hline 
$\vartheta_{0^{<}}$  & $-2.6208\times10^{-4}$  & $-2.6208\times10^{-4}$  & $-2.6208\times10^{-4}$  & $-2.6208\times10^{-4}$\tabularnewline
\hline 
$\Delta\vartheta_{+1^{<}}$  & $-11.5719$  & $-9.6609$  & $-8.3838$  & $-7.4221$\tabularnewline
$\Delta\vartheta_{+2^{<}}$  & $-1.2227$  & $-8.0648\times10^{-1}$  & $-6.1750\times10^{-1}$  & $-5.0363\times10^{-1}$\tabularnewline
$\Delta\vartheta_{+3^{<}}$  & $-5.6756\times10^{-2}$  & $-3.6433\times10^{-2}$  & $-2.7577\times10^{-2}$  & $-2.2344\times10^{-2}$\tabularnewline
\hline 
$\Delta\vartheta_{-1^{<}}$  & $11.5724$  & $9.6614$  & $8.3843$  & $7.4225$\tabularnewline
$\Delta\vartheta_{-2^{<}}$  & $1.2228$  & $8.0655\times10^{-1}$  & $6.1756\times10^{-1}$  & $5.0368\times10^{-1}$\tabularnewline
$\Delta\vartheta_{-3^{<}}$  & $5.6761\times10^{-2}$  & $3.6436\times10^{-2}$  & $2.7580\times10^{-2}$  & $2.2346\times10^{-2}$\tabularnewline
\hline 
$\mu_{0^{<}}$  & $2.7475\times10^{-21}$  & $2.7475\times10^{-21}$  & $2.7475\times10^{-21}$  & $2.7475\times10^{-21}$\tabularnewline
\hline 
$\mu_{+1^{<}}$  & $-6.3652\times10^{-17}$  & $-8.2124\times10^{-17}$  & $-9.0031\times10^{-17}$  & $-9.2892\times10^{-17}$\tabularnewline
$\mu_{+2^{<}}$  & $-2.6383\times10^{-17}$  & $-1.9202\times10^{-17}$  & $-1.5425\times10^{-17}$  & $-1.2950\times10^{-17}$\tabularnewline
$\mu_{+3^{<}}$  & $-1.3787\times10^{-18}$  & $-9.3375\times10^{-19}$  & $-7.2756\times10^{-19}$  & $-6.0023\times10^{-19}$\tabularnewline
\hline 
$\mu_{-1^{<}}$  & $6.3652\times10^{-17}$  & $8.2124\times10^{-17}$  & $9.0031\times10^{-17}$  & $9.2892\times10^{-17}$\tabularnewline
$\mu_{-2^{<}}$  & $2.6383\times10^{-17}$  & $1.9202\times10^{-17}$  & $1.5425\times10^{-17}$  & $1.2950\times10^{-17}$\tabularnewline
$\mu_{-3^{<}}$  & $1.3787\times10^{-18}$  & $9.3375\times10^{-19}$  & $7.2756\times10^{-19}$  & $6.0023\times10^{-19}$\tabularnewline
\hline 
\end{tabular}\caption{The angular position $\vartheta_{\pm n^{<}}$ and the magnification
factor $\mu_{\pm n^{<}}$ of images inside the critical curve for
a point-like source in JNW singularities with varying $\gamma$. The
parameters $M$, $D_{OL}$, $D_{LS}$ and $\beta$ are chosen to be
consistent with those presented in Table \ref{table:out}. All angles
are expressed in units of microarcseconds.}
\label{table:in}
\end{table}

When $b<b_{c}$, photons emitted from the source have the capability
to traverse the singularity, resulting in additional images within
the critical curve. In this scenario, the deflection angle $\alpha$
is computed using eqn. $\left(\ref{eq:IT}\right)$. For light rays
that pass through the singularity without orbiting it, their impact
parameter $b$ is often much smaller than $b_{c}$, leading to
\begin{equation}
\alpha\simeq2b\int_{r_{g}}^{\infty}\frac{dr}{C(r)}=\frac{b}{M}.
\end{equation}
Consequently, the resulting image with $n=0^{<}$ has
\begin{equation}
\vartheta_{0^{<}}\simeq-\frac{D_{OS}M}{D_{LS}D_{OL}}\beta\text{ and }\mu_{0^{<}}^{r}=\mu_{0^{<}}^{t}\simeq-\frac{D_{OS}M}{D_{LS}D_{OL}},
\end{equation}
where we assume $M\ll D_{LS}D_{OL}/D_{OS}$. Notably, the $n=0^{<}$
image is barely distorted by gravitational lensing since $\mu_{0^{<}}^{r}=\mu_{0^{<}}^{t}$.

Relativistic images produced by light rays traversing the singularity
have been discussed in a generic spherically symmetric metric using
the strong deflection approximation \cite{Chen:2023trn}. Applying
these calculations to the JNW singularity, we obtain
\begin{equation}
\alpha=-\bar{a}^{<}\ln\left(b^{2}/b_{c}^{2}-1\right)+\bar{b}^{<}+\mathcal{O}\left(\left(b^{2}/b_{c}^{2}-1\right)\ln\left(b^{2}/b_{c}^{2}-1\right)\right),
\end{equation}
where
\[
\bar{a}^{<}=2\text{ and }\bar{b}^{<}=2\ln\left[\frac{4\left(2\gamma+1\right)}{2\gamma-1}\right]+I_{R}^{<}.
\]
As a result, the angular position of the $n^{\text{th}}$-order images
is given by
\begin{equation}
\vartheta_{\pm n^{<}}=\vartheta_{\pm n^{<}}^{0}-\frac{b_{c}e_{n}^{<}D_{OS}}{2\bar{a}^{<}D_{LS}D_{OL}}\frac{\left(\beta-\vartheta_{\pm n^{<}}^{0}\right)}{\left(1+e_{n}^{<}\right){}^{3/2}},\label{eq:theta-in}
\end{equation}
where
\begin{equation}
\vartheta_{\pm n^{<}}^{0}=\pm\frac{b_{c}}{D_{OL}}\frac{1}{\sqrt{1+e_{n}^{<}}}\text{ and }e_{n}^{<}=e^{\frac{\bar{b}^{<}-2\pi n}{\bar{a}^{<}}}.
\end{equation}
The corresponding radial and tangential magnification factors are
\begin{equation}
\mu_{\pm n^{<}}^{r}=-\frac{b_{c}e_{n}^{<}D_{OS}}{2\bar{a}^{<}D_{OL}D_{LS}}\left(1+e_{n}^{<}\right)^{2/3}\text{ and }\mu_{\pm n^{<}}^{t}=\pm\frac{b_{c}}{\beta D_{OL}\sqrt{1+e_{n}^{<}}},
\end{equation}
respectively, showing that the relativistic images are highly stretched
along the critical curve.

Similarly, in the aforementioned astrophysical scenario, Table \ref{table:in}
presents the angular position and magnification factor of the images
inside the critical curve. For the $n=0^{<}$ image, its angular position
and magnification factor hardly depend on $\gamma$. It shows that
the $n=0^{<}$ image is almost centered in the image plane, positioned
far away from the critical curve. Moreover, compared to the relativistic
images listed in the table, the $n=0^{<}$ image has a smaller magnification
factor, mainly due to the very small value of $\mu_{0^{<}}^{t}$.
Notably, the relativistic images inside the critical curve are more
widely separated and magnified compared to the images outside the
critical curve, resulting from the significant bending of light rays
upon entering or exiting the photon sphere. This bending also leads
to a noticeable dependence of the angular positions and magnification
factors on $\gamma$. As $\gamma$ increases, the relativistic images
tend to approach the critical curve. Additionally, there is a marked
rise (fall) in the magnification factor of the $n=\pm1^{<}$ ($n=\pm2^{<}$
and $\pm3^{<}$) images with the increase of $\gamma$.

\section{Conclusions}

\label{sec:Con}

This paper investigated the phenomenon of gravitational lensing by
JNW naked singularities, which possess a photon sphere. Similar to
Schwarzschild black holes, light rays that orbit the photon sphere
outside of it result in multiple images of a distant source beyond
the critical curve. However, when photons originating from the source
enter the photon sphere, they are observed to approach the singularity
in a finite coordinate time. Assuming that the singularity is remedied
by a regular core, these photons can traverse the regularized singularity,
leading to the emergence of new images inside the critical curve.
In particular, the $n=0^{<}$ image, formed by light rays passing
directly through the singularity, remains well-separated from the
critical curve and exhibits minimal distortion. Furthermore, we have
demonstrated that relativistic images inside the critical curve are
more magnified and positioned farther from the critical curve than
those outside the critical curve, enhancing the possibility of resolving
relativistic images within the critical curve. As a result, these
findings present a powerful means of detecting and studying JNW naked
singularities through their distinct gravitational lensing signatures.

Although current observational facilities lack the capability to distinguish
new images within the critical curve in JNW singularity spacetime,
the next-generation Very Long Baseline Interferometry has emerged
as a promising tool for this purpose \cite{Johnson:2019ljv,Himwich:2020msm,Gralla:2020srx}.
Hence, it would be highly intriguing to extend our analysis to encompass
more astrophysically realistic models, such as the rotating JNW naked
singularity solution and the imaging of accretion disks around JNW
singularities.

Lastly, it is important to note that the optical appearances of distant
sources within the JNW singularity spacetime are contingent upon the
specific regularization method applied to address the singularity.
In cases where the singularity is regularized by a wormhole throat
instead of a regular core, photons reaching the throat will traverse
to another universe. In scenarios where the source and observer inhabit
the same universe, the resultant images are solely located outside
the critical curve, bearing a close resemblance to the black hole
case. In contrast, if the source and the observer reside in different
universes, only images situated within the critical curve become observable,
thereby presenting observational phenomena distinct from those associated
with black holes.
\begin{acknowledgments}
We are grateful to Qingyu Gan and Xin Jiang for useful discussions
and valuable comments. This work is supported in part by NSFC (Grant
No. 12105191, 12275183, 12275184 and 11875196). Houwen Wu is supported
by the International Visiting Program for Excellent Young Scholars
of Sichuan University. 
\end{acknowledgments}

 \bibliographystyle{unsrturl}
\bibliography{ref}

\end{document}